\documentclass[preprint,amsmath,amssymb]{revtex4}
\usepackage{mathrsfs}
\usepackage{epsfig,bm,dcolumn}
\usepackage{graphicx}
\usepackage{color}
\usepackage{amsmath}

\begin{document}

\title{Ginzburg-Landau theory of a trapped Fermi gas with a BEC-BCS crossover}

\author{Kun Huang, Zeng-Qiang Yu, Lan Yin}
\email{yinlan@pku.edu.cn}
\address{School of Physics, Peking University, Beijing 100871,
China}

\date{\today}

\begin{abstract}
The Ginzburg-Landau theory of a trapped Fermi gas with a BEC-BCS
crossover is derived by the path-integral method.  In addition to
the standard Ginzburg-Landau equation, a second equation describing
the total atom density is obtained.  These two coupled equations are
necessary to describe both homogeneous and inhomogeneous systems.
The Ginzburg-Landau theory is valid near the transition temperature
$T_c$ on both sides of the crossover.  In the weakly-interacting BEC
region, it is also accurate at zero temperature where the
Ginzburg-Landau equation can be mapped onto the Gross-Pitaevskii
(GP) equation.  The applicability of GP equation at finite
temperature is discussed. On the BEC side, the fluctuation of the
order parameter is studied and the renormalization to the molecule
coupling constant is obtained.
\end{abstract} 

\maketitle

\section{Introduction}\label{Chap1}

In a Fermi gas with a BEC-BCS crossover, the scattering length $a_s$
can be tuned by the technique of Feshbach resonance \cite{Feshbach}.
The effective interaction between atoms is proportional to the
scattering length.  A dilute Fermi gas with a negative scattering
length is in a BCS pairing state below a critical temperature
\cite{Jin, Ketterle04, Thomas, Chin2}, very similar to BCS
superconductors.  A dilute Fermi gas with a positive scattering
length is in a BEC state of diatomic molecules below another
critical temperature \cite{Jochim, Greiner}.   Although the
scattering length is divergent at the resonance, the system evolves
smoothly between the BCS state and the molecular BEC state across
the resonance at low temperatures. The observation of the BEC-BCS
crossover \cite{Ketterle05} provided a new platform to study
strong-correlation effects in fermionic systems.

The BEC-BCS crossover can be qualitatively understood in the
BCS-type mean-field theory\cite{leggett1980, eagles1969}.  In this
theory, as the interaction changes sign from attractive to repulsive
across the resonance, the pair size of Cooper pairs decreases, and
eventually these atom pairs become diatomic molecules. Although the
mean-field theory offers the correct physical picture, it
overestimates the critical temperature and the molecule-molecule
scattering length in the weakly-interacting BEC limit.  Nozi\`{e}res
and Schmitt-Rink (NSR) \cite{NSR1985} found that fluctuation effects
have to be considered to get the correct critical temperature.  The
total fermion density includes not only the mean-field fermion
density but also the density of thermal molecules, which provides
the important relation between the density and the chemical
potential. The NSR theory is essentially equivalent to treating
Gaussian fluctuations in the Ginzburg-Landau theory
\cite{Randeria1993}.  The NSR theory was also applied at zero
temperature and the molecule-molecule scattering length was found in
good agreement \cite{HuHui2006, Randeria2008} with the few-body
calculation \cite{Petrov}.

The purpose of this paper is to construct the Ginzburg-Landau theory
to describe the BEC-BCS crossover in a trapped Fermi gas.  Compared
to microscopic theories, the Ginzburg-Landau theory has potential
advantages of requiring less computation and being easier to be
applied to inhomogeneous cases such as trapped systems.  In the
weakly-interacting BEC region, the Ginzburg-Landau equation was
shown to be equivalent to the Gross-Pitaevskii (GP) equation at zero
temperature \cite{GP-BdG}. In the unitary region, a modified
Ginzburg-Landau theory was developed to describe the phase slip
\cite{yin1}, vortex \cite{yin2}, and vortex lattices \cite{yin3}.
However there still lacks a complete Ginzburg-Landau description of
the whole BEC-BCS crossover.  In the following, we first derive the
Ginzburg-Landau theory of a trapped Fermi gas by the
functional-integral method, and obtain the Ginzburg-Landau equation
and the equation for the fermion density. The density equation is
important for providing the density profile of the BEC-BCS crossover
in both the inhomogeneous and homogeneous cases. Then we concentrate
on weakly-interacting BEC limit, study the Ginzburg-Landau equation
at both zero and finite temperatures, and consider effects due to
fluctuations of the order parameter.  The conclusion is given in the
end.


\section{Ginzburg-Landau theory of a trapped Fermi gas}\label{Chap2}

A Fermi gas with a wide Feshbach resonance can be effectively
described by a single-channel model, while for the narrow resonance
case a two-channel model is more accurate \cite{Strinati}.  In this
paper we consider only the wide resonance case in which the
single-channel Hamiltonian density is given by \begin{equation}
{\mathscr
H}(x)=\sum_\sigma\phi^{\dag}_\sigma(x)\bigg[-\frac{\hbar^2\nabla^2}{2m}+V({\bf
r})\bigg]\phi_\sigma(x) +g
\phi^{\dag}_{\uparrow}(x)\phi^{\dag}_{\downarrow}(x)\phi_{\downarrow}(x)\phi_{\uparrow}(x),
\label{H} \end{equation} where $x=({\bf r}, \tau)$ is the coordinate
in space and time, $\phi_\sigma(x)$ is the field operator of atoms
with spin-component $\sigma$, $m$ is the mass of a Fermi atom,
$V({\bf r})$ is the external trapping potential, and the coupling
constant is given by $g=4\pi\hbar^2a_s/m$.  In the following we
consider only the spin-balanced case where the densities of spin-up
and spin-down atoms are the same.

The grand partition function $Z$ can be written in
functional-integral formalism as \begin{equation} Z=\int{\mathscr
D}\phi^*_\sigma{\mathscr D}\phi_\sigma \exp[S_\phi],
\end{equation} with the action given by\begin{equation}
 S_\phi=\int{\mathrm d}^4x\bigg\{-\sum_\sigma \phi^*_\sigma(x)(\partial_\tau-\mu)\phi_{\sigma}(x)-{\mathscr H}[\phi^*,\phi]\bigg\}
\end{equation} where  $0\leq\tau\leq1/(k_BT)$, and $\mu$ is the
chemical potential.

The interaction term in Eq. (\ref{H}) can be decoupled by
introducing an auxiliary field $\Delta(x)$ and applying the
Hubbard-Stratanovich transformation \cite{Randeria1993}.  After
integrating out the fermion field $\phi(x)$, we obtain
\begin{eqnarray} Z&=&\int{\mathscr D}\Delta^*{\mathscr D}\Delta \exp[S_\Delta], \end{eqnarray} where the
action $S_\Delta$ is given in terms of the auxiliary field $\Delta$,
\begin{equation} S_\Delta=\ln\det{\bf M}+{1\over g}\int{\mathrm
d}^4x|\Delta(x)|^2, \label{pf}\end{equation}
\begin{equation}
{\bf M}=\bigg[\begin{array}{cc}-\partial_\tau-\hat{H}({\bf
r})&-\Delta(x)\\
-\Delta^*(x)&-\partial_\tau+\hat{H}({\bf r})\end{array}\bigg],
\end{equation}
and
\begin{equation}\hat{H}({\bf
r})=-\frac{\hbar^2\nabla^2}{2m}+V({\bf
r})-\mu.\label{h}\end{equation} The action $S_\Delta$ can be further
separated into two parts,
\begin{equation}
S_\Delta=S_{eff}+S_0,
\end{equation}
where $S_0= \ln\det{\bf M_0}$ is independent of $\Delta$ with
\begin{equation} {\bf
M}_0=\bigg[\begin{array}{cc}-\partial_\tau-\hat{H}({\bf
r})&0\\
0&-\partial_\tau+\hat{H}({\bf r})\end{array}\bigg], \end{equation}
and the second part $S_{eff}$ vanishes when $\Delta=0$,
\begin{eqnarray} S_{eff} &=&\frac{1}{g}\int{\mathrm d}^4x|\Delta(x)|^2+{\rm
Tr}\ln[{\bf I}-{\bf G}{\bf u}], \end{eqnarray} with ${\bf u}={\bf M}_0-{\bf M}=\bigg[\begin{array}{cc}0&\Delta(x)\\
\Delta^*(x)&0\end{array}\bigg]$, ${\bf I}$ as the identity
matrix, and ${\bf G}={\bf M}^{-1}_0=\bigg[\begin{array}{cc}G_+&0\\
0&G_-\end{array}\bigg]$ being the Green's function of a
noninteracting Fermi gas.

Next we expand the second term in $S_{eff}$ to the fourth order in
$\Delta$. This approximation holds only when $\Delta$ is small
compared to the Fermi energy, which is true when $|T-T_c|/T_c\ll1$
or the system is in the weakly-interacting BEC regime. After the
expansion, we obtain
\begin{eqnarray}
&&S_{eff}\approx\frac{1}{g}\int{\mathrm d}^4x|\Delta(x)|^2+
\int{\mathrm d}^4x{\mathrm
d}^4x_1Q(x,x_1)\Delta^*(x)\Delta(x_1)\nonumber\\&&-\frac{1}{2}\int
\prod_{i=1}^4{\mathrm d}^4x_i \, R(x_1,...,x_4)
\Delta^*(x_1)\Delta(x_2)\Delta^*(x_3)\Delta(x_4),\nonumber\\
\label{Seff_raw}
\end{eqnarray}
where
\begin{eqnarray} Q(x_1,x_2)&=&-G_+(x_1,x_2)G_-(x_2,x_1),\nonumber\\
R(x_1,...,x_4)&=&G_+(x_1,x_2)G_-(x_2,x_3)G_+(x_3,x_4)G_-(x_4,x_1).\nonumber\\\end{eqnarray}

Since we are interested in low-energy and long-wavelength properties
of the system, we apply gradient expansion in Eq. (\ref{Seff_raw})
and obtain
\begin{eqnarray}\label{gdexp}
S_{eff}&\approx&\int{\mathrm
d}^4x\bigg[\,d\,\Delta^*(x)\partial_\tau\Delta(x)+c\,\Delta^*(x)\frac{\hbar^2\nabla^2}{4m}\Delta(x)\nonumber\\&&+(a+\frac{1}{g})|\Delta(x)|^2-\frac{1}{2}b|\Delta(x)|^4\bigg],
\end{eqnarray}
where
\begin{eqnarray}
a&=& \int{\mathrm d}^4x'Q(x-x'/2,x+x'/2),\nonumber\\
b&=&\int \prod_{i=1}^3{\mathrm d}^4x_i R(x,x_1,x_2,x_3),\nonumber\\
c&=&\frac{4m}{\hbar^2}\int{\mathrm d}^4x'\frac{{r'}^2}{6}
Q(x-x'/2,x+x'/2),\label{co_raw}
\end{eqnarray}
and $x'=({\bf r}',\tau')$.  The coefficient $d$ of the
time-derivative term is defined as the coefficient of the linear
term in the expansion of $Q$ in the frequency space in the
zero-frequency limit,
\begin{equation}
d=\lim_{\omega\rightarrow 0}\int{\mathrm
d}^4x'{e^{i\omega\tau'}-1\over i\omega}
Q(x-x'/2,x+x'/2).\end{equation} The equation of motion of the order
parameter can be obtained by taking $\delta
S_{eff}/\delta\Delta^*(x)=0$, which yields
\begin{equation}\label{motion}
(d\,\partial_\tau+c\,\frac{\hbar^2\nabla^2}{4m}+a+\frac{1}{g})\Delta(x)
-b|\Delta(x)|^2\Delta(x)=0.
\end{equation}

For an arbitrary trap potential $V({\bf r})$, it is difficult to
obtain the exact analytical expression of the Green's function ${\bf
G}$. Here we consider only the case where the trap length is much
larger than the inter-particle distance, and the local-density
approximation (LDA) can be applied,
\begin{eqnarray} {\bf G}(x_1,x_2)\approx {\bf
G}^{(0)}(x_1,x_2).\end{eqnarray}
Here ${\bf G}^{(0)}=\bigg[\begin{array}{cc}G_+^{(0)}&0\\
0&G_-^{(0)}\end{array}\bigg]$ is the Green's function of a
homogeneous noninteracting Fermi gas with the chemical potential
$\mu'=\mu-V[({\bf r}_1+{\bf r}_2)/2]$,
\begin{equation}\bigg[\begin{array}{cc}-\partial_\tau-\hat{T}+\mu'&0\\
0&-\partial_\tau+\hat{T}-\mu'\end{array}\bigg]{\bf
G}^{(0)}(x-x^\prime)=\delta(x-x^\prime){\bf I}\label{G0},
\end{equation} where $\hat{T}=-\hbar^2\nabla^2/(2m)$.
In LDA, the coefficients in Eq.(\ref{co_raw}) are given by
\begin{eqnarray} a&=&\int{{\mathrm d}^3k \over (2\pi)^3}\frac{\tanh[\epsilon_{\bf k}/(2k_BT)]}{2\epsilon_{\bf k}},\nonumber\\
 b&=&\int{{\mathrm d}^3k \over (2\pi)^3}\bigg[\frac{\tanh[\epsilon_{\bf k}/(2k_BT)]}{4\epsilon_{\bf k}^3}-
\frac{\mathrm{sech}^2[\epsilon_{\bf k}/(2k_BT)]}{8k_BT\epsilon_{\bf k}^2}\bigg],\nonumber\\
c&=&\int{{\mathrm d}^3k \over (2\pi)^3}\bigg[
\frac{\tanh[\epsilon_{\bf k}/(2k_BT)]}{4\epsilon_{\bf k}^2} -\frac{
\mathrm{sech}^2[\epsilon_{\bf k}/(2k_BT)]}{8k_BT\epsilon_{\bf
k}}\bigg],\nonumber\\
d&=&\int{{\mathrm d}^3k \over (2\pi)^3} \frac{\tanh[\epsilon_{\bf
k}/(2k_BT)]}{4\epsilon^2_{\bf k}},\label{Dl0_equation}
\end{eqnarray}
where $\epsilon_{\bf k}=\hbar^2k^2/(2m)-\mu'$. It is important to
note that in obtaining the coefficients given by Eq.
(\ref{Dl0_equation}) LDA is applied to the Green's function of a
noninteracting Fermi gas ${\bf G}^{(0)}$, which is different from
applying LDA directly to the broken symmetry state.  The latter case
requires that the gap has to be much larger than the trap frequency,
which does not hold near the critical temperature or in the
weakly-interacting BEC regime.

In the unitary and BCS regime where there is a clear Fermi surface,
the Fermion energy $\epsilon_{\bf k}$ has zero points, and there are
divergences in integrands on r.h.s. of Eq. (\ref{Dl0_equation}) at
zero temperature, suggesting that this formalism does not work at
zero temperature.  At finite temperatures, the integrand on r.h.s.
of the equation for the coefficient $d$ also diverges.  A more
careful treatment of this coefficient leads to damping in the
dynamics of the order parameter \cite{Popov}. In the following, we
will not study the dynamics or zero-temperature properties in this
regime so these issues do not occur.

At the stationary state, the function $\Delta({\bf r})$ satisfy the
saddle-point condition
\begin{equation} {\delta S_{eff}\over \delta \Delta({\bf r})}=0,
\end{equation}
which leads to the Ginzburg-Landau equation
\begin{equation}
-c\frac{\hbar^2}{4m}\nabla^2\Delta({\bf
r})-(a+\frac{1}{g})\Delta({\bf r}) +b|\Delta({\bf r})|^2\Delta({\bf
r})=0.
\end{equation}
The Ginzburg-Landau equation can be rewritten in terms of a
wave-function $\psi({\bf r})$ defined by $\psi({\bf r}) \equiv
\sqrt{c}\Delta({\bf r})$,
\begin{equation}
-\frac{\hbar^2}{4m}\nabla^2\psi({\bf r})+\alpha\psi({\bf r})+
\beta|\psi({\bf r})|^2\psi({\bf r})=0, \label{order_eq}
\end{equation}
where
\begin{eqnarray}
\alpha&=&-{1 \over c}(a+{1 \over g}), \nonumber \\
\beta&=&{b \over c^2}, \label{ab}
\end{eqnarray}
and the gradient terms of $c$ are ignored due to its small spatial
variation within LDA.  In Eq. (\ref{ab}), the coupling constant is
given by
\begin{equation}
g^{-1}={m \over 4 \pi\hbar^2 a_s}-\int{{\mathrm d}^3 k \over (2\pi)^3}{1
\over 2\epsilon_{\bf k}^0},
\end{equation}
where $\epsilon_{\bf k}^0=\hbar^2k^2/(2m)$, and the second r.h.s. term is a
counter term in the particle-particle channel.

In the simple homogeneous case, the critical temperature $T_c$ is
determined from the equation
\begin{equation}
\alpha=0,
\end{equation}
which is exactly the $T_c$ equation in the mean-field BCS theory,
\begin{equation}\int{{\mathrm d}^3k \over (2\pi)^3}\bigg\{\frac{\tanh[\epsilon_{\bf k}/(2k_BT_c)]}{2\epsilon_{\bf k}}
-\frac{1}{2\epsilon_{\bf k}^0}\bigg\}=-\frac{m}{4\pi\hbar^2a_s}.
\end{equation}
Above $T_c$, the
Ginzburg-Landau equation does not have a nontrivial solution. Close
to $T_c$, to the first order of $T-T_c$, the coefficient $\alpha$ is
approximately given by
\begin{equation}
\alpha\approx\lambda(T_c)(T-T_c), \end{equation} where
$$\lambda(T_c)=\int{{\mathrm d}^3k \over (2\pi)^3}{1\over 4c k_BT_c^2}{\rm sech}^2[{\epsilon_{\bf
k} \over 2k_BT_c}].$$ In the trapped case, the critical temperature
$T_c$ is approximately determined by
\begin{equation}
\alpha({\bf r_0})=0,
\end{equation}
where ${\bf r_0}$ is the place with the highest fermion density.

The Ginzburg-Landau equation determines the distribution of the
order parameter in the stationary state.  To obtain the density
distribution, we need to study the thermodynamic potential $\Omega$,
given by
\begin{equation}\Omega=-k_BT\ln Z=\Omega_f+\Omega_s,
\end{equation}
where $\Omega_f$ is thermodynamic potential of a noninteracting
Fermi gas, in LDA it is given by
\begin{equation}
\Omega_f=2\int {\mathrm d}^3r\int {{\mathrm d}^3k \over
(2\pi)^3}[\epsilon_{\bf k}+k_BT\ln n_{\bf k}],
\end{equation}
and $n_{\bf k}=1/\{1+\exp[\epsilon_{\bf k}/(k_BT)]\}$ is the Fermi
distribution function.  The term $\Omega_s$ is the contribution to
thermodynamic potential by the order parameter.  If we consider only
the stationary state and ignore fluctuations of the order parameter
in the mean-field approximation, $\Omega_s$ is approximately given
by
\begin{eqnarray}\Omega_s^{(0)}&=&-k_BT\,S_{eff}^{(0)}\nonumber\\
&=&\int{\rm d}^3r\left[-\psi^*({\bf r})\frac{\hbar^2}
{4m}\nabla^2\psi({\bf r})+ \alpha|\psi({\bf r})|^2+\frac{
\beta}{2}|\psi({\bf
r})|^4\right].\label{S0}\nonumber\\
\end{eqnarray}
From the relation $N=-\partial\Omega/\partial\mu$, the atom density
in the mean-field approximation can be obtained,
\begin{eqnarray}
n({\bf r}) \approx n_f({\bf r})+n_s^{(0)}({\bf r}),\label{n_MFT}
\end{eqnarray}
where $n_f$ associated with $\Omega_f$ is the density of a
homogeneous noninteracting Fermi gas with the chemical potential
$\mu'=\mu-V({\bf r})$ ,
\begin{equation}
n_f({\bf r})=2\int{{\mathrm d}^3k \over (2\pi)^3} n_{\bf k}({\bf
r}),
\end{equation}
and $n_s^{(0)}$ associated with $\Omega_s^{(0)}$ is the density due
to the order parameter,
\begin{equation}
n_s^{(0)}({\bf r})\approx\frac{\partial a}{\partial\mu} |\Delta({\bf
r})|^2=2|\psi({\bf r})|^2\label{n_0_raw}.
\end{equation}
Note that in Eq. (\ref{n_0_raw}) the quartic term in $\psi({\bf r})$
is ignored because it is much smaller than the quadratic term. The
density equation (\ref{n_MFT}) indicates that the total density can
be separated into two parts, i.e. $n_f$ from the normal state and
$n_s^{(0)}$ from the superfluid order parameter, which is consistent
with the two-fluid model of a superfluid at finite temperatures. The
superfluid atom pairs are described by the wave-function $\psi({\bf
r})$.  In a trapped system, usually the total number of atoms $N$ is
given,
\begin{equation}
N=\int {\mathrm d}^3 r n({\bf r}),
\end{equation}
from which the chemical potential $\mu$ can be solved.  The
Ginzburg-Landau equation (\ref{order_eq}) and the density equation
(\ref{n_MFT}) provide a complete mean-field phenomenological
description of a trapped superfluid Fermi gas.

\section{Ginzburg-Landau equation in the weakly interacting BEC regime}

In the weakly interacting BEC regime, the Fermi gas is dilute,
$na_s^3\ll 1$, and the order parameter $\Delta$ is much less than
the binding energy of a diatomic molecule given by
$\epsilon_0=\hbar^2/(ma_s^2)$.  In this regime, since the chemical
potential $\mu$ is negative, there is no Fermi surface.  The
coefficients in the Ginzburg-Landau equation given by Eq.
(\ref{Dl0_equation}) are well defined even at zero temperature.
Therefore the Ginzburg-Landau theory can be applied from zero
temperature to near $T_c$.  In this section we study the
Ginzburg-Landau equation in this regime.

At zero temperature, from Eq. (\ref{Dl0_equation}) and (\ref{ab}),
the coefficients in the Ginzburg-Landau equation are given by
\begin{eqnarray}
\alpha_0&=&4\sqrt{|\mu'|}\big(\sqrt{|\mu'|}-\sqrt{\epsilon_0/2}\big), \nonumber\\
\beta_0&=&\frac{2\sqrt{2}\pi\hbar^3}{m^{3/2}\sqrt{|\mu'|}},
\nonumber \\
d_0&=&c_0. \label{ztcf}
\end{eqnarray}
Since the density $n_0=|\alpha_0/\beta_0|$ is much smaller than
$1/a_s^3$, we obtain $\mu'= -(\epsilon_0/2)[1+{\rm O}(n_0 a_s^3)]$,
and approximately
\begin{eqnarray}
\alpha_0&\approx&-(2\mu'+\epsilon_0)\nonumber,\\
\beta_0&\approx&\frac{8\pi\hbar^2a_s}{2m}.
\end{eqnarray}
With the time-dependent term, the Ginzburg-Landau equation can be
written as
\begin{equation}\label{zteq}
\partial_\tau\psi(x)-\frac{\hbar^2\nabla^2}{4m}\psi(x)+[2V({\bf
r})-\mu_b]\psi(x)+\frac{8\pi\hbar^2a_s}{2m} |\psi(x)|^2\psi(x)=0,
\end{equation}
where $\mu_b=2\mu+\epsilon_0$ is the chemical potential of
molecules. This time-dependent Ginzburg-Landau equation is identical
to the Gross-Pitaevskii equation of molecular BEC if the imaginary
time $\tau$ is analytically continued to the real time $t$, $\tau=i
\hbar t$, with $m_b=2m$ identified as the molecule mass and
$a_b=2a_s$ identified as the scattering length between molecules.
The density equation (\ref{n_MFT}) in this regime is trivial,
$n=2|\psi|^2$, which means all the atoms are paired into condensed
molecules at zero temperature.  However, the molecule scattering
length $a_b$ extracted from Eq. (\ref{zteq}) is $a_b=2a_s$,
contradicting to the result $a_b\approx 0.6a_s$ from the few-body
calculation \cite{Petrov}.  This discrepancy is due to the fact that
we have not considered the fluctuation effect which are discussed in
the next section.

In the weakly interacting regime, the BEC transition temperature
$T_c$ is much smaller than the molecule binding energy $\epsilon_0$,
$k_BT_c\ll\epsilon_0$.  As a result, at any finite temperature below
or near $T_c$, the coefficients in the time-dependent
Ginzburg-Landau equation are almost the same as those at zero
temperature given by Eq. (\ref{zteq}), except the molecular chemical
potential $\mu_b$ now varying with temperature.  Therefore the
time-dependent Ginzburg-Landau equation (\ref{zteq}) remains valid
at finite temperatures below or near $T_c$.


\section{Fluctuation effects}

In this section, we consider the effect due to the fluctuation of
the order parameter which is ignored in our derivation so far. The
effective action in terms of the pair wave-function
$\psi(x)=\sqrt{c}\Delta(x)$ is given by
\begin{eqnarray} \label{seff} S_{eff}&=&\int{\rm
d}^4x\bigg[\kappa\,\psi^*(x)\partial_\tau\psi(x)+\psi^*(x)\frac{\hbar^2}
{4m}\nabla^2\psi(x)\nonumber\\&&-
\alpha|\psi(x)|^2-\frac{1}{2}\beta|\psi(x)|^4\bigg],\end{eqnarray}
where $\kappa=d/c$.  For simplicity we consider only the homogeneous
case. In the following, we concentrate on the weakly-interacting BEC
regime near or below $T_c$, where $\kappa \approx 1$, $\alpha
\approx -\mu_b$, and $\beta \approx 8\pi\hbar^2a_s/(2m)$.  In this
regime, the effective action given by Eq. (\ref{seff}) is the same
action of a Bose gas with the boson mass given by $2m$ and a
scattering length given by $2a_s$.

The effective action given by Eq. (\ref{seff}) provides a
contribution to the thermodynamic potential given by
\begin{equation}\Omega_s=
-k_BT\ln\int\mathscr{D}\psi^*\mathscr{D}\psi\,\,\exp[S_{eff}],\label{Omg2}
\end{equation} from which we can obtain its contribution to the density $n_s$ by taking
$-\partial\Omega_s/\partial \mu$. For a weakly-interacting Bose gas,
Bogoliubov's theory is accurate \cite{Fetter}, in which $n_s$ is
given by
\begin{equation}
n_s \approx 2\psi_0^2+\int {{\mathrm d}^3 k \over
(2\pi)^3}[{(\epsilon_{b{\bf k}}-\alpha)\over E_{b{\bf
k}}}\coth({E_{b{\bf k}}\over 2k_BT})-1],
\end{equation}
where $\epsilon_{b{\bf k}}=\hbar^2k^2/(4m)$ is the kinetic energy of
a molecule, and $E_{b{\bf k}}=\sqrt{\epsilon_{b{\bf
k}}(\epsilon_{b{\bf k}}-2\alpha)}$ is the excitation energy of the
molecular quasi-particle.  The order parameter $\psi_0$ is
equivalent to the expectation value of the Bose field-operator,
$\psi_0=\sqrt{-\alpha/ \beta}$, where its phase is chosen so
$\psi_0$ is positive for simplicity.

In Bogoliubov's theory of a dilute Bose gas, there is an
ultra-violet divergence appearing in the calculation of the
thermodynamical potential $\Omega_s$ and the ground state energy,
which is removed by the renormalization of the coupling constant
$$g_b^{-1}={2m \over 4 \pi\hbar^2 a_b}-\int{{\mathrm d}^3 k \over (2\pi)^3}{1
\over 2\epsilon_{b{\bf k}}},$$ where the second r.h.s. term is a
counter term only appearing in the calculation involving the
particle-particle channel.  However in the effective action Eq.
(\ref{seff}), the constant $\beta$ is not renormalized so far, which
would result in a divergent term in $\Omega_s$ and the ground state
energy \cite{Fetter2} given by
\begin{equation}\label{t1p}
\Omega_s'=-\alpha^2 \sum_{\bf k}{1 \over 4\epsilon_{b{\bf k}}}.
\end{equation}
with the contribution to density given by
\begin{equation}\label{t2p}
n_s'=-\alpha \int {{\mathrm d}^3k \over (2\pi)^3}{1 \over
\epsilon_{b{\bf k}}},
\end{equation}
where $\partial \alpha /\partial \mu \approx -2$.  This ultra-violet
divergence is an unphysical result, due to the invalidity of the
gradient expansion used in derivation the Ginzburg-Landau action
given by Eq. (\ref{gdexp}) at short distances.  Fluctuations inside
molecules become important at short distances, which is beyond the
description of the gradient expansion.  The energy scale at which
the gradient expansion is invalid is approximately given by the
molecule binding energy $\epsilon_0=\hbar^2/(ma_s^2)$, which is
equal to the molecular kinetic energy at wavevector $k=2/a_s$. Thus
a straightforward renormalization method is to put a cutoff
$\Lambda=2/a_s$ in the $k-$integrals of Eq. (\ref{t1p}) and
(\ref{t2p}), which yields
\begin{equation}
n_s'=-{4\alpha m \over \pi^2 \hbar^2 a_s}={16 \over \pi} \psi_0^2.
\end{equation}
After considering the renormalization due to fluctuations, we obtain
the density equation,
\begin{widetext}
\begin{eqnarray}
n&=&n_f+n_s+n_s'=2(1+{8 \over \pi}) \psi_0^2+\int{{\mathrm d}^3 k
\over (2\pi)^3}[{(\epsilon_{b{\bf k}}-\alpha)\over  E_{b{\bf
k}}}\coth({E_{b{\bf k}}\over 2k_BT})-\tanh({\epsilon_{\bf k}\over
2k_BT})].
\end{eqnarray}
\end{widetext}

At zero temperature, the density equation is simply given by
\begin{eqnarray} n&=&2(1+{8 \over \pi})\psi_0^2+\int{{\mathrm d}^3 k \over (2\pi)^3}[{(\epsilon_{b{\bf k}}-\alpha)\over  E_{b{\bf
k}}}-1] \nonumber \\
&=& 2(1+{8 \over \pi})\psi_0^2+{16\over 3}\psi_0^3\sqrt{{8 \over
\pi}a_s^3}. \label{n_zero}
\end{eqnarray} The first r.h.s term in Eq. (\ref{n_zero}) implies
that the condensate density is $n_0=2(1+8/\pi)\psi_0^2$ not simply
$2\psi_0^2$.  The second r.h.s term is proportional to $\psi_0^3$,
which comes from the quantum depletion of molecules. Thus in term of
the true molecular condensate wave-function
\begin{equation} \tilde{\psi}(x)=\sqrt{(1+8/\pi)}\psi(x),\label{wave}\end{equation}
the Ginzburg-Landau equation given by Eq.(\ref{zteq}) should be
rewritten as
\begin{widetext}
\begin{equation}
\partial_\tau\tilde{\psi}(x)-\frac{\hbar^2\nabla^2}{4m}\tilde{\psi}(x)
+[2V({\bf r})-\mu_b]\tilde{\psi}(x)
+\frac{4\pi\hbar^2a_b}{2m}|\tilde{\psi}(x)|^2\tilde{\psi}(x)=0,
\label{GP_eq}
\end{equation}
\end{widetext} where $a_b$ is the scattering length of molecules after considering the
renormalization,
\begin{equation} \label{reab}
a_b={ 2a_s \over 1+8/ \pi}\approx0.56a_s,
\end{equation}
very close to the result $a_b\approx0.6a_s$ from the few-body
calculation \cite{Petrov}.

\begin{figure}
\includegraphics[width=4in]{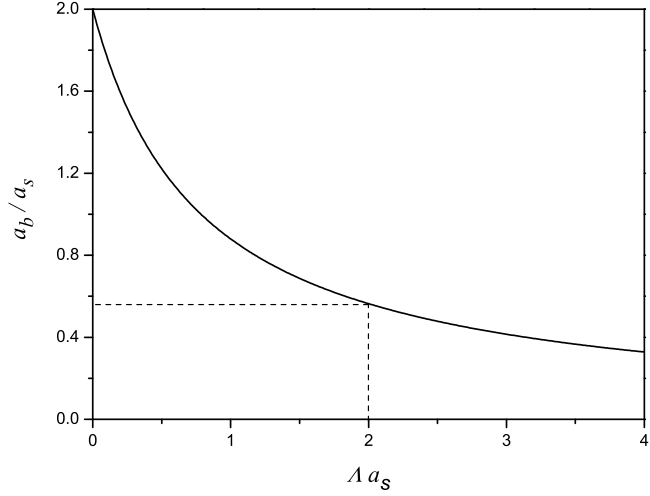}
\caption{The molecule scattering length $a_b$ vs the cutoff
$\Lambda$. The dotted lines are $\Lambda=2/a_s$, and $a_b= 0.56 a_s$
which is very close to the result $a_b\approx0.6a_s$ from the
few-body calculation \cite{Petrov}.  At $\Lambda=0$, the mean-field
result $a_b=2a_s$ is recovered.} \label{fig_ab}
\end{figure}

Although the renormalization to the mean-field molecule scattering
length given by Eq. (\ref{reab}) was obtained below $T_c$, it is
valid above $T_c$ as well, which can be understood in the vacuum
renormalization of the molecule coupling constant in the $T$-matrix
approximation,
\begin{eqnarray}
g_b^{-1}&=&g_{b0}^{-1}-i\int{{\mathrm d}\omega \over
2\pi}\int{{\mathrm d}^3k \over (2\pi)^3}G_b({\bf k},\omega)G_b(-{\bf
k},-\omega)\nonumber\\
&=&g_{b0}^{-1}+\int{{\mathrm d}^3k \over (2\pi)^3}{1\over
2\epsilon_{b{\bf k}}}, \label{gb}
\end{eqnarray}
where $g_{b0}=8\pi\hbar^2a_s/(2m)$ is the molecule coupling constant
in the mean-field approximation, $g_b=4\pi\hbar^2a_b/(2m)$ is the
renormalized coupling constant, and $$G_b({\bf k},\omega)={1 \over
\omega-\epsilon_{b{\bf k}}+i\delta}$$ is the Green's function of a
molecule in vacuum.  A cutoff $\Lambda=2/a_s$ should be put in the
$k-$integral of Eq. (\ref{gb}) due to the same reason as stated
above that this effective description cease to be accurate when
$\epsilon_{b{\bf k}}\geq\epsilon_0$, leading to the same
renormalization,
\begin{equation}
a_b={2a_s \over 1+4a_s\Lambda/\pi},
\end{equation}
same as Eq. (\ref{reab}) at $\Lambda=2/a_s$.   The molecule
scattering length $a_b$ as a function of the cutoff $\Lambda$ is
shown in Fig. \ref{fig_ab}.  It should be emphasized that a more
precise cutoff than $2/a_s$ is necessary for any further numerical
comparison with the few-body result, which requires understanding of
high-energy processes beyond the Ginzburg-Landau description.  The
vacuum renormalization of the molecule coupling constant can also be
applied to trapped systems within LDA as long as the molecule
binding energy is much bigger than trap frequencies.

At $T_c$, $\alpha=0$, the density is given by
\begin{equation}
n=2\int{{\mathrm d}^3 k \over (2\pi)^3}\{{1 \over
\exp[\epsilon_{b{\bf k}}/(k_BT_c)]-1}+{1\over \exp[\epsilon_{\bf
k}/(k_BT_c)]+1}\},
\end{equation}
indicating that there are only thermally-excited atoms and
molecules.  Since $k_BT_c \ll \epsilon_0$, the number of
thermally-excited atoms are negligiblly small, and almost all the
particles are thermal molecules,
\begin{equation}
n \approx \int{{\mathrm d}^3 k \over (2\pi)^3}{2 \over
\exp[\epsilon_{b{\bf k}}/(k_BT_c)]-1}.
\end{equation}

\begin{figure}
\includegraphics[width=4in]{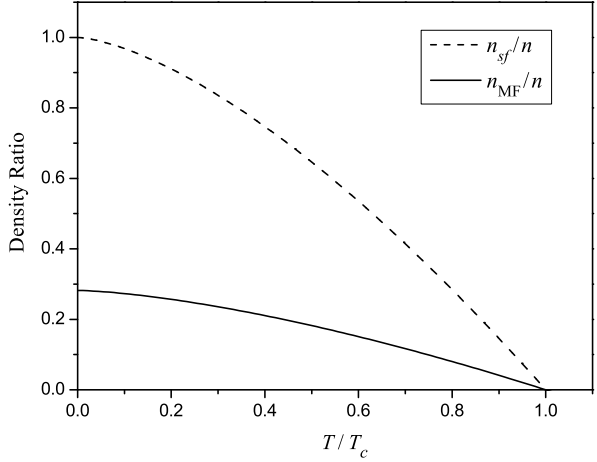}
\caption{Density ratios of a homogeneous Fermi gas in the
weakly-interacting BEC limit as functions of temperature.  The solid
line is the mean-field density given by Eq.(\ref{n_MFT}) divided by
the total density.  For comparison, the dashed line is the ratio of
the superfluid density to the total density.} \label{fig}
\end{figure}

Generally at finite temperature below $T_c$, the density can be
separated into the superfluid density $n_{sf}$ and the normal
density $n_n$, $n=n_{sf}+n_n$.  The superfluid density is given by
the condensate density and the quantum depletion,
\begin{equation}
n_{sf}=2\tilde{\psi}_0^2\left(1+{8\over 3}\sqrt{\tilde{\psi}_0^2
a_b^3 \over \pi}\right),
\end{equation}
consistent with traditional theories of a dilute Bose gas
\cite{Fetter}. The normal density is given by the total density of
thermal atoms and molecules. Since the thermal atoms are negligible,
the normal density is approximately given by the density of thermal
molecules,
\begin{equation}
n_n=2\int{{\mathrm d}^3 k \over (2\pi)^3}{(\epsilon_{b{\bf
k}}-\alpha)\over E_{b{\bf k}}}{1 \over \exp[E_{b{\bf k}}/(k_BT)]-1}.
\end{equation}
Compared with the mean-field result in Eq (\ref{n_MFT}), both the
superfluid density and the normal density are renormalized.  In the
weakly-interacting BEC limit, as shown in Fig. \ref{fig}, the
renormalization to the density is quite strong.  At zero
temperature, the mean-field density account for only about 28\% of
the total density, while the rest density is due to fluctuation
contribution. At $T_c$, almost all the density is due to fluctuation
contribution.

Away from the weakly-interacting BEC limit, the fluctuation effect
is more difficult to deal with in the Ginzburg-Landau theory.  The
coefficients in the Ginzburg-Landau equation are more complicated
than those given by Eq. (\ref{ztcf}), and fluctuations of the order
parameter cannot be simply treated by the Bogoliubov's theory.
Moreover the wavevector cutoff $2/a_s$ vanishes in the unitary
region where a more subtle renormalization scheme is required. These
problems will be explored in our future work.  The situation is
simpler again on the other side, in the weakly-interacting BCS
limit, where the fluctuation of the order parameter is strongly
damped.  In this limit, the mean-field theory is accurate and the
fluctuation of the order parameter is less important.

\section{Conclusion}

In conclusion, we have derived the Ginzburg-Landau theory of a
trapped Fermi gas with a BEC-BCS crossover.   Two equations
including the standard Ginzburg-Landau equation and the density
equation are obtained to describe the order parameter distribution
and the density profile. In the weakly-interacting BEC limit, the
Ginzburg-Landau equation is equivalent to the Gross-Pitaevskii (GP)
equation.  The fluctuation of the order parameter is strong in this
limit, which can be treated by the Bogoliubov's theory. Compared
with mean-field results, both the density and the molecule-molecule
scattering length are renormalized, in agreement with the few-body
and NSR theories.  This work is supported by NSFC under Grant No.
10674007, and by Chinese MOST under grant number 2006CB921402.


\begin{thebibliography}{99}
\bibitem{Fetter2}
The divergence is in the first r.h.s term of Eq. (35.25) in Ref.
\cite{Fetter}.
\end{thebibliography}
\end{document}